\documentclass[floats,prl,twocolumn]{revtex4}%
\usepackage{amsfonts}
\usepackage{amsmath}
\usepackage{amssymb}
\usepackage{graphicx}%
\providecommand{\U}[1]{\protect\rule{.1in}{.1in}}
\providecommand{\U}[1]{\protect\rule{.1in}{.1in}}
\providecommand{\U}[1]{\protect\rule{.1in}{.1in}}
\providecommand{\U}[1]{\protect\rule{.1in}{.1in}}
\providecommand{\U}[1]{\protect\rule{.1in}{.1in}}
\providecommand{\U}[1]{\protect\rule{.1in}{.1in}}
\providecommand{\U}[1]{\protect\rule{.1in}{.1in}}
\providecommand{\U}[1]{\protect\rule{.1in}{.1in}}

\begin{document}
\title{Comment on Generating Unexpected Spin Echoes in Dipolar Solids with $\pi$ Pulses.}
\author{}
\author{Mar\'{\i}a Bel\'{e}n Franzoni}
\affiliation{FaMAF, Universidad Nacional de C\'{o}rdoba, 5000, C\'{o}rdoba, Argentina}
\author{Claudia M. S\'{a}nchez}
\affiliation{FaMAF, Universidad Nacional de C\'{o}rdoba, 5000, C\'{o}rdoba, Argentina}
\author{Patricia R. Levstein}
\email{patricia@famaf.unc.edu.ar}
\affiliation{FaMAF, Universidad Nacional de C\'{o}rdoba, 5000, C\'{o}rdoba, Argentina}
\startpage{1}
\maketitle

The authors of \cite{barrett07} reported NMR measurements in solids with
different pulse sequences. In some multiple $\pi$\ pulse echo trains they
observed \ \emph{\textquotedblleft coherent\textquotedblright} signals of
several seconds, well beyond the Hahn echo decay, $T_{2HE}$ (few
milliseconds). They ascribed the long tails to coherent phenomena arising on
the dipolar couplings during any real finite NMR pulse. Their statement is
based on assumptions and parameters in the calculations, that do not represent
the experimental conditions: While the $\pi$ pulse has a typical duration of
12.5 $\mu$s, the separation\ between pulses in the train is extremely
shortened (1 $\mu$s) with respect to the experimental values ($\approx$ 1 ms)
and the dipolar coupling was multiplied by a factor 25 \cite{barrettarXiv},
for the calculation in Fig 3 of \cite{barrett07}. Under these conditions, the
dipolar evolution$\mathbb{\ }$during the pulses becomes relevant compared to
any evolution during the interpulse separations of 1 $\mu$s. Moreover, in
their calculations the Zeeman term is simplified by setting $\omega_{z_{i}%
}=\Omega_{z}$ for every spin $i$, after assuming that $\omega_{z_{i}}$ is
extremely uniform. At this time, our respective groups recognize that in the
samples manifesting these effects, the Zeeman line broadening dominates the
spectrum%
\'{}%
s full width at half maximum  \cite{belpat,belpat2,barrettprl08}.

We have proposed \cite{belpat} that the long tails are a consequence of the
slowness of the spin diffusion in between pulses, together with an rf field
inhomogeneity or an inhomogeneous line that enables non-perfect $\pi$\ pulses
across the sample. \ A detailed measurement of the tails as a function of the
interpulse time, $\tau,$ was performed \cite{belpat2}. The CPMG1 echoes decay
with a short time, associated to $T_{2HE\text{ \ }}$and a long time, $t_{l}$,
that characterizes the tails (Fig.5 in \cite{belpat2}). Here, a plot of the
amplitudes $A_{s}$\ and $A_{l}$\ obtained from the fittings as a function of
$\tau$ is displayed. The amplitude $A_{\emph{l}}$\ is a quantification of the
tail magnitude. Both plots show that the tails decay much faster and their
magnitudes decrease, tending to disappear, as $\tau$\ becomes longer. This
supports the idea that once the flip-flop becomes operative between pulses,
the small components of magnetization leading to the formation of stimulated
echoes averages out \cite{belpat}.\begin{figure}[ptb]
\begin{center}
\includegraphics[
natheight=3.298400in,
natwidth=4.401000in,
height=2.6775in,
width=3.5483in
]{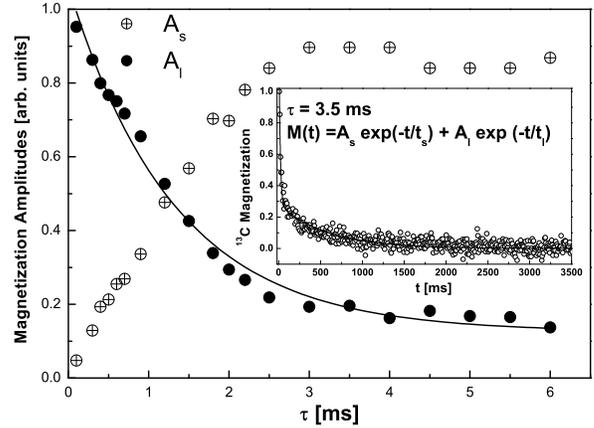}
\end{center}
\caption{Amplitudes of a double exponential fitting of the CPMG1 echoes fixing
$t_{\emph{s}}=T_{\emph{2HE}},$ as a function of $\tau$. }%
\label{amplitudes}%
\end{figure}Moreover, the most revealing experiment confirming that the
dipolar evolution \emph{during} the pulses is not relevant, consists of
studying the dependence of the tails as a function of pulse duration. An
experiment varying the rf field strengths up to values above 400 times the
full width at half maximum, FWHM, was performed by Li et al (Fig.13 in
\cite{barrettarXiv}). The results show a remarkable insensitivity of the long
tails to the rf field strength for values above four times the FWHM. \ This
independence of the tails on $t_{p}$\ supports our statements. Summarizing,
the common characteristic of all the samples that present long tails is that
the Zeeman shift distribution is much larger than the dipolar interaction,
regardless of the isotopic dilution or the origin of the inhomogeneity. The
long tails are a marker of the absence of spin diffusion.

\end{document}